 \documentclass[final,5p,times,twocolumn]{elsarticle}

\usepackage{amssymb}
\usepackage{lipsum}
\usepackage{booktabs}
\usepackage{multirow}
\usepackage{xurl}
\usepackage{hyperref}
\hypersetup{
    colorlinks=true,
    linkcolor=blue,
    filecolor=magenta,      
    urlcolor=cyan,
    citecolor=blue,
    }

\newcommand{\kms}{km\,s$^{-1}$}

\begin{document}

\begin{frontmatter}



\title{Effect of event classifiers on jet quenching-like signatures in high-multiplicity $p+p$ collisions at $\sqrt{s} = 13$ TeV}

\author[first]{Hushnud Hushnud}
\ead{hushnud.hushnud@cern.ch}
\affiliation[first]{organization={University Centre of Research and Development Department, Chandigarh University, },
            addressline={Gharuan}, 
            city={Mohali},
            postcode={140413}, 
            state={Punjab},
            country={India}}

\author[second]{Omveer Singh}
\affiliation[second]{organization={Institut f$\ddot{u}$r Kernphysik, Goethe-Universit$\ddot{a}$t},
            addressline={Max-von-Laue-Str. 1}, 
            city={Frankfurt am Main},
            postcode={60438}, 
            country={Germany}}

\author[third]{Srikanta Kumar Tripathy}
\affiliation[third]{organization = {Warsaw University of Technology,
            Faculty of Physics, Nuclear Physics Division},
            addressline={Koszykowa 75},
            city={Warsaw},
            postcode={00-662},
            country={Poland}}

\author[fourth]{Aditya Nath Mishra}
\affiliation[fourth]{organization={University Centre of Research and Development Department, Chandigarh University, },
            addressline={Gharuan}, 
            city={Mohali},
            postcode={140413}, 
            state={Punjab},
            country={India}}
            
\author[corresponding]{Kalyan Dey}
\ead{kalyn.dey@gmail.com (corresponding author)}
\affiliation[corresponding]{organization={Department of Physics, Bodoland University},
        addressline={Rangalikhata}, 
            city={Kokrajhar},
            postcode={783370}, 
            state={Assam},
            country={India}}
            

\begin{abstract}
The motivation behind exploring jet quenching-like phenomena in small systems arises from the experimental observation of heavy-ion-like behavior of particle production in high-multiplicity proton-proton ($p+p$) collisions. The study of jet quenching in $p+p$ collisions can help to resolve the ongoing debate about the underlying cause of heavy-ion-like collective behavior observed in small systems where the probability of production of quark-gluon plasma (QGP) is believed to be negligible. However, quantifying the jet quenching in $p+p$ collisions is a challenging task, as the magnitude of the nuclear modification factor ($R_{\rm AA}$ or $R_{\rm CP}$), which is used to quantify jet quenching, is influenced by several factors, such as the estimation of centrality and the scaling factor. The most common method of centrality estimation employed by the ALICE collaboration is based on measuring charged-particle multiplicity with the V0 detector situated at the forward rapidity. This technique of centrality estimation makes the event sample biased towards hard processes like multijet final states. This bias of the V0 detector towards hard processes makes it difficult to study the jet quenching effect in high-multiplicity $p+p$ collisions. In the present article, we propose to explore the use of a new and robust event classifier, flattenicity which is sensitive to both the multiple soft partonic interactions and hard processes. The $\mathcal{P}_{\rm CP}$, a quantity analogous to $R_{\rm CP}$, has been estimated for high-multiplicity $p+p$ collisions at $\sqrt{s} = 13$ TeV using \texttt{PYTHIA8} model for both the V0M (the multiplicity classes selected based on V0 detector acceptance) as well as flattenicity. The evolution of $\mathcal{P}_{\rm CP}$ with $p_{\rm T}$ shows a heavy-ion-like effect for flattencity which is attributed to the selection of softer transverse momentum particles in high-multiplicity $p+p$ collisions.  
\end{abstract}



\begin{keyword}
Jet quenching \sep V0M \sep Flattenicity\sep \texttt{PYTHIA} \sep High-multiplicity



\end{keyword}

\end{frontmatter}




\section{Introduction}\label{introduction}
\label{intro}

The Large Hadron Collider (LHC) is an excellent experimental facility where the nucleus-nucleus ($A+A$) collisions at relativistic energies are carried out for studying conditions similar to those found in the early universe, where a noble phase of matter known as quark-gluon plasma (QGP) is believed to have existed. The observation of a large value of elliptic flow ($v_2$) and a perfect NCQ (number of constituent quarks) scaling of $v_2$ at the RHIC beam energies suggested the formation of an almost perfect fluid with color degrees of freedom \cite{Adamczyk_2013}. The first evidence of jet quenching was observed at the RHIC in the early 2000s \cite{Drees_2002}. This provides direct evidence of the formation of a strongly interacting hot and dense medium in high-energy heavy-ion collisions. 
Recently, the study of high-multiplicity (HM) $p+p$ collisions at LHC energies has taken center stage in the investigation of QGP-like scenarios owing to the revelation of some surprising observations such as strangeness enhancement \cite{nature:2017} and the ridge-like structure in two-particle azimuthal correlation \cite{Khachatryan_2010, PhysRevLett.Khachatryan:2016, Velicanu:2011, PhysRevLett.Aad:2016} in HM $p+p$ collisions. The jet quenching is believed to be the most promising signature that can be studied in small systems in order to shed some light on the origin of heavy-ion-like features seen in HM $p+p$ collisions. One way to measure jet quenching is via measuring the nuclear modification factor ($R_{\rm AA}$ or $R_{\rm CP}$) as a function of transverse momentum. $R_{\rm AA}$ is defined as the ratio of the yield of particles produced in $A+A$ collisions to that of $p+p$ collisions, scaled by the number of binary nucleon-nucleon collisions $\langle N_{\rm coll} \rangle$. In contrast, $R_{\rm{CP}}$ is defined as the ratio of the yield of the particles produced in the central to that of the peripheral $A+A$ collisions, where both the numerators and denominators are scaled by the $\langle N_{\rm coll} \rangle$ in respective centrality bins. Experimentally, the value of this scaling factor, i.e. $\langle N_{\rm coll} \rangle$ in the case of heavy-ion collisions is estimated with the help of the Glauber model \cite{Miller:2017}. However, since the Glauber model assumes the colliding objects to be extended nuclei with a finite size and density distribution, for single-nucleon projectiles such as protons, the Glauber model is not at all suitable. The ambiguity in measuring the scaling factor in the case of $p+p$ collisions makes it challenging to measure jet quenching in small systems in terms of the nuclear modification factor. The measurement of jet quenching is further complicated by the difficulty in defining centrality in $p+p$ collisions.
The centrality or event activity in $p+p$ collisions is, therefore, measured by the absolute number of particles produced in an event. The most common experimental method for determining event activity in $p+p$ collisions is to count the total number of charged particles produced at a given rapidity.
However, this method is unsuitable if the observable of interest is also measured from the same rapidity window due to the presence of inherent auto-correlations \cite{PhysRevC.99.024906:auto}. The other most widely used event activity estimation technique, usually employed by the ALICE collaboration, utilizes the total charge deposited in the forward V0 detectors. In this technique, the charge distribution is divided into various multiplicity classes known as V0M. Though this method is free from auto-correlations, but it is biased towards hard collisions. This method, therefore, is expected to influence various measured quantities, such as transverse momentum distribution, jet yield, etc., and precautions need to be taken while interpreting different results based on V0M. Various event-shape observables such as transverse sphericity ($S_{\rm T}$) \cite{Abelev_2012}, transverse spherocity ($S_{0}$) \cite{Acharya_2019, Ortiz_2015} and the relative transverse activity classifier ($R_{\rm T}$) \cite{Martin_2016, Ortiz_2017} have also been used to measure event activity in $p+p$ collisions. These event classifiers also suffer from biases originating due to hard gluon radiation \cite{Ortiz_2023}. Flattenicity is the newest addition to the race for measuring event activity in $p+p$ collisions, as proposed by Ortiz \textit{et al.} \cite{ortiz2022look}. This technique is sensitive to both multi-partonic soft interactions (MPI) and multijet final states (hard collisions) \cite{Ortiz_2023}. In the present contribution, an attempt has been made to study the jet quenching-like signatures in $p+p$ collisions in terms of a variable ($\mathcal{P}_{\rm CP}$) proposed in Ref. \cite{Dey:2022} which is analogous to the nuclear modification factor ($R_{\rm CP}$) in $A+A$ collisions. In this work, the quantity $\mathcal{P}_{\rm CP}$ is first estimated using the new event classifier flattenicity and compared with the standard V0M technique. \\~\\
\vspace{-1pt}
 The paper is structured as follows. In Section \ref{PYTHIA8}, a brief account of the \texttt{PYTHIA8} event generator is provided. In Section \ref{Analysis details}, the formalism to select the multiplicity classes by employing the V0M and the flattenicity event classifier is discussed. The $\mathcal{P}_{\rm CP}$ as a function of transverse momentum for both the event classifiers is presented in Section \ref{Result}. Finally, we have summarized the important outcomes of the present analysis in Section \ref{Summary}.  

\section{PYTHIA8 model}\label{PYTHIA8}
The \texttt{PYTHIA8} \cite{Sj_strand_2006, Sj_strand_2008, Sj_strand_2015, 10.21468/SciPostPhysCodeb.8} is a general-purpose Monte Carlo event generator that is widely used to simulate ultra-relativistic collisions among leptons, protons, and nuclei. A wide range of physics processes such as hard and soft scattering, initial and final state parton showers, multi-parton interactions (MPI), fragmentation, color reconnection, and decay processes are incorporated in the \texttt{PYTHIA8} model. The composite nature of the proton leads to multiple parton-parton interactions in a single event which is one of the key aspects of the \texttt{PYTHIA8} model \cite{Sj_strand_2006}. This MPI feature of \texttt{PYTHIA8} very well describes various experimental results available for small systems \cite{Abazov_2010, Chekanov_2008}. However, the \texttt{PYTHIA8} with MPI was not able to describe the transverse momentum distribution \cite{Adam_2016}, multiplicity-dependent production of strange particles \cite{Abelev_2012_strange, nature:2017} and jet production rate \cite{Acharya_2019, Acharya_2022} at LHC energies. Hence, a new final state phenomenon, known as the color reconnection (CR) mechanism \cite{Argyropoulos_2014}, is introduced in \texttt{PYTHIA8} model. \\~\\  
\vspace{-1pt}
In the CR mechanism, two partons produced from independent hard scattering are color reconnected and make a large transverse boost, which increases with the number of MPIs \cite{gustafson2009}. In the MPI-based CR model \cite{Skands_2014} of \texttt{PYTHIA8}, the parton at the low-$p_{\rm T}$ MPI system connects with the partons at the high-$p_{\rm T}$ MPI system. The incorporation of CR in \texttt{PYTHIA8} model successfully explains many experimental features such as multiplicity dependence of mean transverse momentum \cite{Abelev_2013} and the linear increase of J/$\psi$ with multiplicity at forward rapidities \cite{Abelev_2012_Jpsi, Thakur_2018} in $p+p$ collisions.
 The CR mechanism is seen to mimic the heavy-ion-like effect in $p + p$ collisions \cite{Ortiz_Velasquez_2013, Dey:2022, Bierlich_2015}. 
A new variant of the CR mechanism known as QCD-based CR \cite{Bierlich_2015} is also included in the  \texttt{PYTHIA8} model where the strings are color reconnected based on the QCD color rules leading to a minimum string length. This new scheme introduces various tunable parameters that qualitatively describe the baryon-to-meson ratio in charm sector in $p+p$ collisions at LHC energies \cite{Acharya_2021, Acharya_2022_baryon}.

\section{Analysis details}\label{Analysis details}
In the present analysis, three different tunings of \texttt{PYTHIA 8.306}, namely, the MPI-based CR, the QCD-based CR, and the no-CR scenario have been used. All the \texttt{PYTHIA8} parameters were set to Monash tune, except for the no-CR mode. The inelastic non-diffractive $p+p$ collisions at $\sqrt{s} = 13$ TeV for each configuration have been generated. The events satisfying the condition of at least one charged particle in the mid-rapidity ($|\eta|$ $<$ 1) window have been selected. Two different event selection classifiers, such as V0M and a new event classifier known as flattenicity, employed in the present investigation are discussed below.
\subsection{V0M}
In the ALICE experiment, the multiplicity classes are determined by using two forward V0 detectors. In the current analysis, the same acceptance coverage as the V0 detector of the ALICE is used. The charged particles are chosen in the pseudorapidity coverage of V0A ($2.8 < \eta < 5.1$) and V0C ($-3.7 < \eta < -1.7$). The total charged particle distribution in V0 acceptance is divided into various multiplicity classes (V0M). The significance of the V0M event estimator is discussed in Section \ref{intro}. 
\subsection{Flattenicity}
Recently, a new event classifier, known as flattenicity, has been proposed by Ortiz \textit{et al.} \cite{ortiz2022look} which is found to be sensitive to both hard collisions as well as the multiple soft partonic interactions \cite{Ortiz_2023}.
In this paper, we investigate the viability of estimating the multiplicity classes in $p+p$ collisions at LHC energies in terms of flattenicity using the acceptance coverage of existing detectors of the ALICE experiment \cite{ALICE_2014} and investigate its potential impact on the jet quenching-like signatures in small systems. In the present investigation, for estimating flattenicity, we followed the same methodology as employed in Ref. \cite{Ortiz_2023}.
Flattenicity is measured on an event-by-event basis by using the particle multiplicity recorded in the pseudorapidity coverage of V0A and V0C and the azimuthal coverage, $0 < \phi < 2\pi$. An event-by-event grid in $\eta$-$\phi$ space is constructed, which is further divided into 64 elementary cells. For each event, the flattenicity is defined as 

\begin{equation}
\rho = \frac{\sqrt{\sum_{i = 0}^{N_{\rm cell}}(N_{\rm ch}^{ \rm{cell},i} - \langle N_{\rm ch}^{\rm cell} \rangle)^{2}/N_{\rm cell}^{2}}}{\langle N_{\rm ch}^{\rm cell} \rangle}
 \label{Eq:Flat}
\end{equation}

\begin{figure}[ht!]
\centering
    \includegraphics[width=\linewidth]{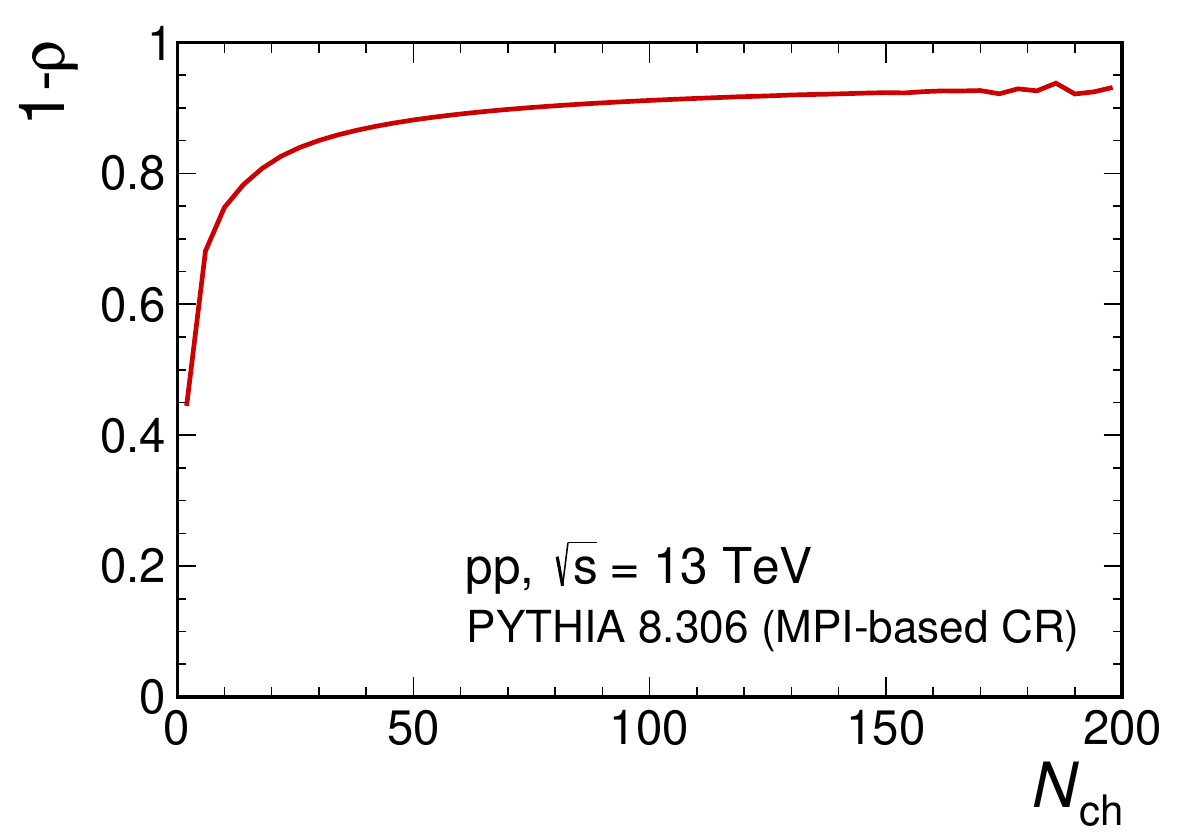}
    \caption{The correlation between the $(1 - \rho)$ and charged-particle multiplicity in the pseudorapidity coverage of the V0 detector for the MPI-based CR model with Monash tune in $p+p$ collision at $\sqrt{s} = 13$ TeV.}
    \label{fig:flat_rho}
\end{figure}

where  $N_{\rm ch}^{\rm{ cell},i}$ is the multiplicity in i$^{th}$ cell, $\langle N_{\rm ch}^{\rm {cell}} \rangle$ is the event-by-event average multiplicity, and $N_{\rm {cell}}$ is the total number of cells. It is seen from Eqn.\ref{Eq:Flat} that the flattenicity is strongly correlated with the event multiplicity as can also be seen from  Fig.~\ref{fig:flat_rho}, where the quantity ($1 - \rho$) is plotted as a function of charged-particle multiplicity in the pseudorapidity coverage of the V0 detector. Fig.~\ref{fig:flat_rho} also reveals that the flattenicity with limit ($1 - \rho$) $\rightarrow 1$ corresponds to high-multiplicity events, whereas low-multiplicity events are associated with ($1 - \rho$) $\rightarrow 0$.

\section{Results and discussion} \label{Result}
\begin{figure}[ht!]   
    \includegraphics[scale = 0.46]{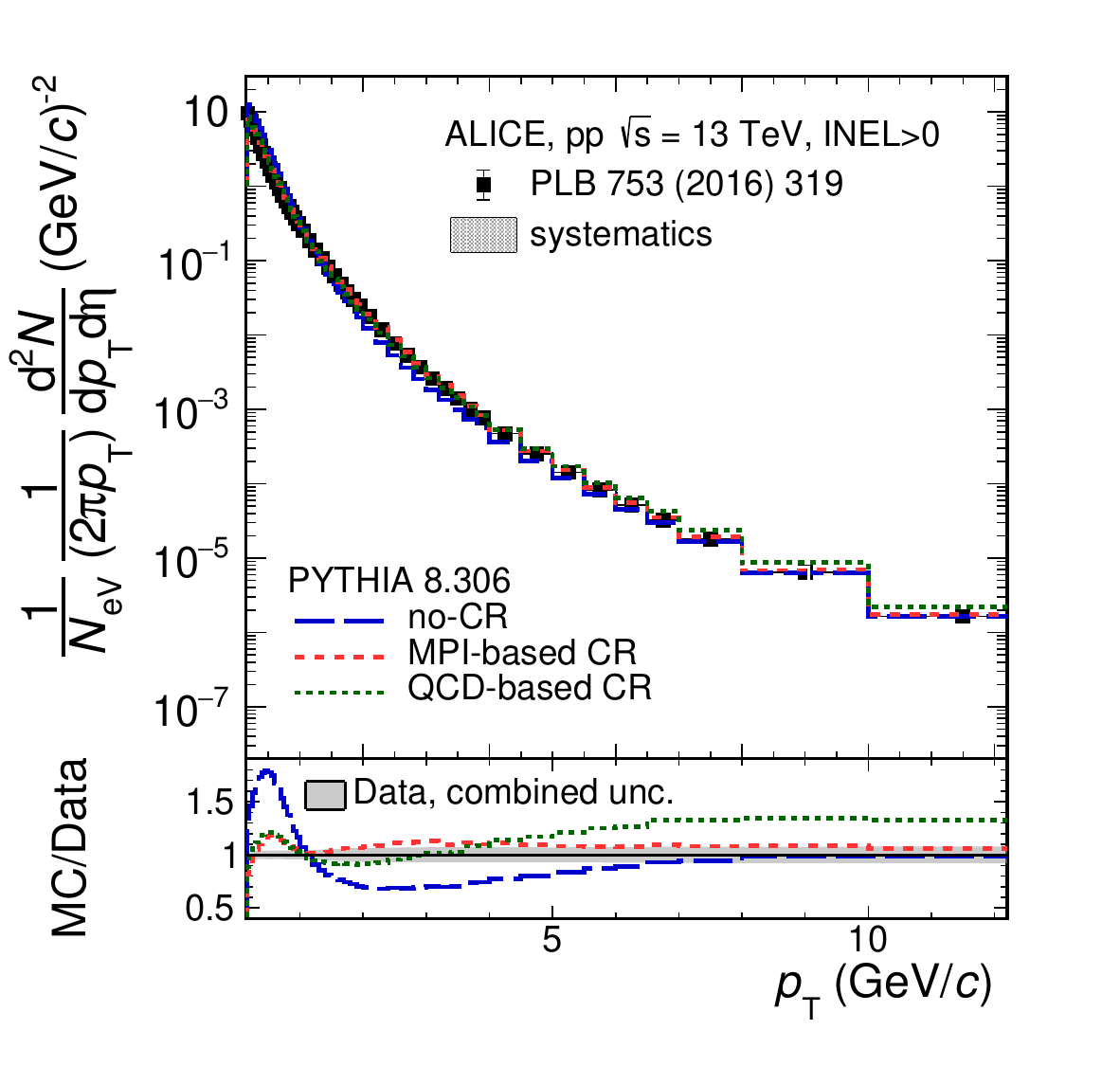}
    \caption{A transverse momentum distribution ($p_{\rm T}$) of the charged particles generated by different variants of \texttt{PYTHIA8} in $p+p$ collisions at $\sqrt{s} = 13$ TeV are compared with the experimental data from the ALICE Collaboration \cite{Adam_2016}. The ratio of models (MC) and data is shown in the lower panel. The systematic uncertainties of the experimental data are displayed as grey bands.}
    \label{fig:pTcharge}
\end{figure}

To begin with, an attempt has been undertaken to select the best possible variant of the \texttt{PYTHIA8} model that can reproduce the transverse momentum spectra of produced charged particles ($\pi^{\pm}$, K$^{\pm}$ and p and $\bar{p}$). For that purpose, apart from MPI-based CR with Monash tune \cite{Skands_2014}, two other variants, such as QCD-based CR and no-CR scenarios have been chosen. The upper panel of Fig.~\ref{fig:pTcharge} shows the transverse momentum distribution ($p_{\rm T}$) of charged particles estimated using different variants of the \texttt{PYTHIA8} model for $p+p$ collisions at $\sqrt{s} = 13$ TeV and compared with the available experimental data from the ALICE Collaboration within the same acceptance \cite{Adam_2016}. On the other hand, the bottom panel of Fig.~\ref{fig:pTcharge} depicts the ratios of different \texttt{PYTHIA8} models with the experimental data. It can be clearly seen from this figure that the default version of \texttt{PYTHIA8} with an MPI-based CR mechanism can give the best possible explanation of the experimental data. The MPI-based CR mechanism of the \texttt{PYTHIA8} reproduces the spectral shape of experimental data within the uncertainties except for $p_{\rm T}$ $<$ $1$ GeV/{\it{c}}. However, the disagreement is not very large and lies within the uncertainties ($\sim$ $20 \%$). The MPI-based CR model is, therefore, chosen for the present analysis. \\~\\

\vspace{-1pt}
The current study is performed by considering the production of charged particles ($\pi^{\pm}$, K$^{\pm}$, $p$ and $\bar{p}$) in the mid-rapidity coverage of $|y|$ $< 0.5$, with- and without using CR tunes of \texttt{PYTHIA8} model. In order to select the multiplicity classes, different event classifiers like V0M and flattenicity are used. \\~\\  
\begin{figure}[h!]
   \centering
   \includegraphics[width=\linewidth]{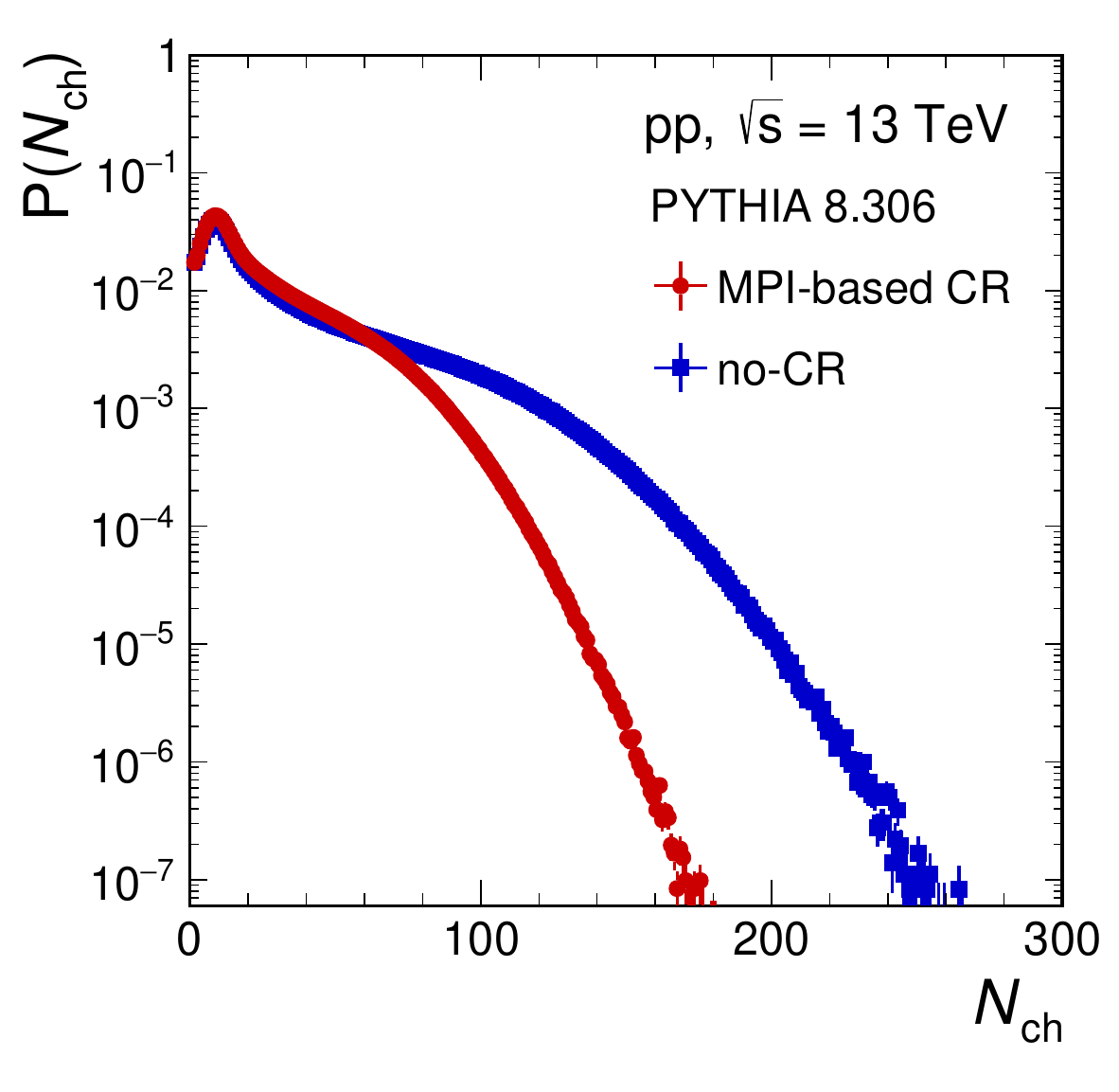}
   \caption{The multiplicity distribution of charged particles using \texttt{PYTHIA8} model with and without the CR mechanism in $p+p$ collisions at $\sqrt{s} = 13$ TeV.}
    \label{fig:charge}
\end{figure}
\vspace{-1pt}
Fig.~\ref{fig:charge} shows the charged particles multiplicity distribution obtained from \texttt{PYTHIA8} for the MPI-based CR with Monash tune and another scenario where the effect of CR is not considered (no-CR). As expected, larger production of charged particles has been observed in the absence of the CR effect as compared to the one with CR mode. The difference in charged-particle multiplicity between CR and no-CR scenarios is due to the merging of different MPIs in the CR mechanism, which leads to a lower production of charged particles \cite{Christiansen_2015}. However, in low multiplicity regions, the effect of CR is very weak, and hence no discernible distinction is seen between the two scenarios. \\~\\

\begin{table*}[htbp!]
\caption{Average charged particle density, $\langle dN_{ch}/d\eta \rangle$, at mid-pseudorapidity ($|\eta|$ $<$ $0.5$) for various multiplicity classes in $p+p$ collisions at $\sqrt{s} = 13$ TeV with different \texttt{PYTHIA8} models. The calculation is performed for different event classifiers.}\label{tab:classes}
\vspace{8pt}
\centering
\large
\begin{tabular}{c|c|c|c|c}
\toprule
\hline
\multirow{3}{5em}{Event \centering Classes} & \multicolumn{4}{c}{$\langle dN_{ch}/d\eta \rangle$} \\
\cline{2-5}
 & \multicolumn{2}{c|} {CR} & \multicolumn{2}{c}{no CR} \\
\cline{2-5} 
& V0M & FL & V0M & FL \\
\hline  
I &23.428 $\pm$ 0.008 & 21.693 $\pm$ 0.009 & 39.744 $\pm$ 0.016 & 36.234 $\pm$ 0.017 \\ 
\hline
II & 18.648 $\pm$ 0.004 & 18.110 $\pm$ 0.004 & 30.759 $\pm$ 0.007 & 29.593 $\pm$ 0.008 \\ 
\hline
III & 15.222 $\pm$ 0.003 & 15.128 $\pm$ 0.003 & 24.003 $\pm$ 0.006 & 23.685 $\pm$ 0.006 \\ 
\hline
IV & 11.736 $\pm$ 0.002 & 11.811 $\pm$ 0.002 & 17.334 $\pm$ 0.004 & 17.406 $\pm$ 0.004 \\ 
\hline
V & 8.515  $\pm$ 0.002 & 8.55 $\pm$ 0.002 & 11.514 $\pm$ 0.003 & 11.802 $\pm$ 0.003 \\ 
\hline
VI & 6.294 $\pm$ 0.001 & 6.319 $\pm$ 0.001 & 7.808 $\pm$ 0.002 & 8.014 $\pm$ 0.002 \\ 
\hline
VII & 4.744 $\pm$ 0.001 & 4.672 $\pm$ 0.001 & 5.545 $\pm$ 0.002 & 5.680 $\pm$ 0.002 \\ 
\hline
VIII & 2.898 $\pm$ 0.001 & 2.925 $\pm$ 0.001 & 3.026 $\pm$ 0.001 & 3.093 $\pm$ 0.001 \\ 
\hline
\bottomrule
\end{tabular}
\end{table*}

\begin{figure}[hb!]
    \centering
    \includegraphics[width=\linewidth]{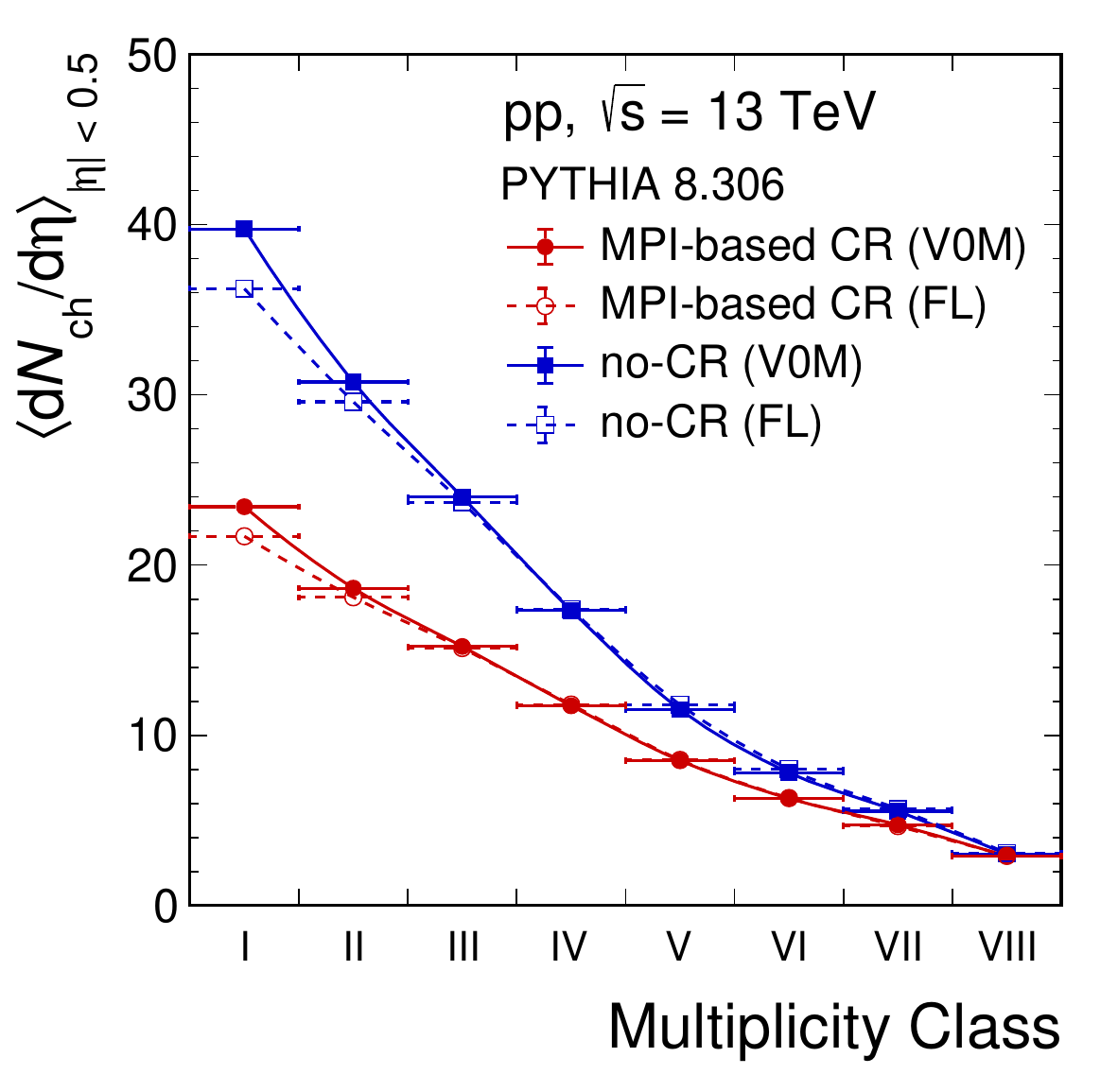}
    \caption{The $\langle dN_{ch}/d\eta \rangle$ in mid-pseudorapidity ($|\eta|<$ $0.5$) as a function of the multiplicity class selected by the different event classifiers for the CR and no-CR scenarios. The full and open markers represent the estimated values of  $\langle dN_{ch}/d\eta \rangle$ using the V0M and flattenicity (FL) estimators, respectively.}
    \label{fig:dnchdeta}
\end{figure}
\vspace{-1pt}

\begin{figure}[ht!]
    \centering
    \includegraphics[width=\linewidth]{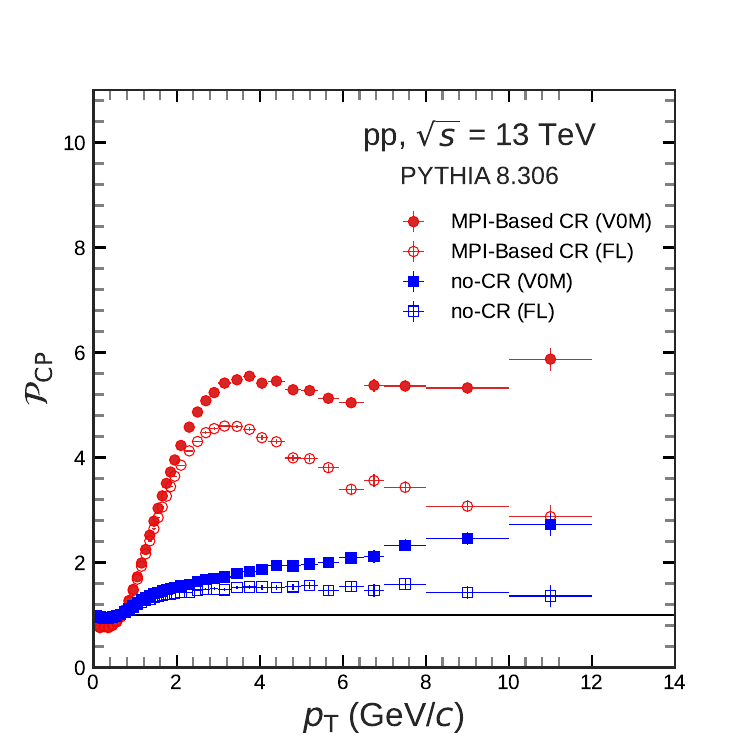}
    \caption{$\mathcal{P}_{\rm CP}$ for charged particles as a function of transverse momentum calculated using the V0M and flattenicity estimators for CR and no-CR scenarios. Full and open markers represent the $\mathcal{P}_{\rm CP}$ values calculated using V0M and flattenicity (FL) estimator respectively.}
    \label{fig:Pcp}
\end{figure}

The charged-particle multiplicity distribution, as shown in Fig.~\ref{fig:charge}, is sliced into various event classes using a standard approach as employed in several experiments at the RHIC and the LHC \cite{Adam_2018, Khachatryan_2017}. This approach of taking the percentile of the charged-particle multiplicity or flattenicity distribution in the pseudorapidity coverage of forward V0 detectors of the ALICE experiment is used for selecting the event classes. The average charged-particle density, $\langle dN_{ch}/d\eta \rangle$, estimated in the mid-pseudorapidity ($|\eta|$ $< 0.5$) for various event classes selected by using V0M and flattenicity for the CR and no-CR scenarios of \texttt{PYTHIA8} are also tabulated in Table~\ref{tab:classes}. \\~\\
\vspace{-1pt}
Fig.~\ref{fig:dnchdeta} displays the charged-particle density estimated in various event classes selected based on flattenicity and V0M multiplicity in $p+p$ collisions at $\sqrt{s}$ = 13 TeV for \texttt{PYTHIA8} with and without CR mechanism.  
Full and open symbols in Fig.~\ref{fig:dnchdeta} represent the $\langle dN_{ch}/d\eta \rangle$ values at mid-pseudorapidity calculated using the flattenicity and V0M multiplicity classes, respectively. 
The larger production of charged particles in the no-CR scenario than the CR mode in HM events as observed in Fig.~\ref{fig:dnchdeta} is also reflected in Fig.~\ref{fig:charge}.
For both the CR and no-CR scenarios, a larger value of $\langle dN_{ch}/d\eta \rangle$ for the high multiplicity bins has been observed when estimated using the V0M estimator than that of the flattenicity approach. The deviation in the value of $\langle dN_{ch}/d\eta \rangle$ for the two studied multiplicity estimators decreases and becomes negligible as one moves towards the lower multiplicity bins. The difference between the two event classifier can be attributed to the fact that, unlike the flattenicity estimator (which is robust against ``hardness" bias), V0M is biased toward harder collisions, resulting in the selection of collisions with high energy-momentum transfer and a higher multiplicity value.\\~\\
\begin{figure*}[ht!]
\centering
   \includegraphics[width=\linewidth]{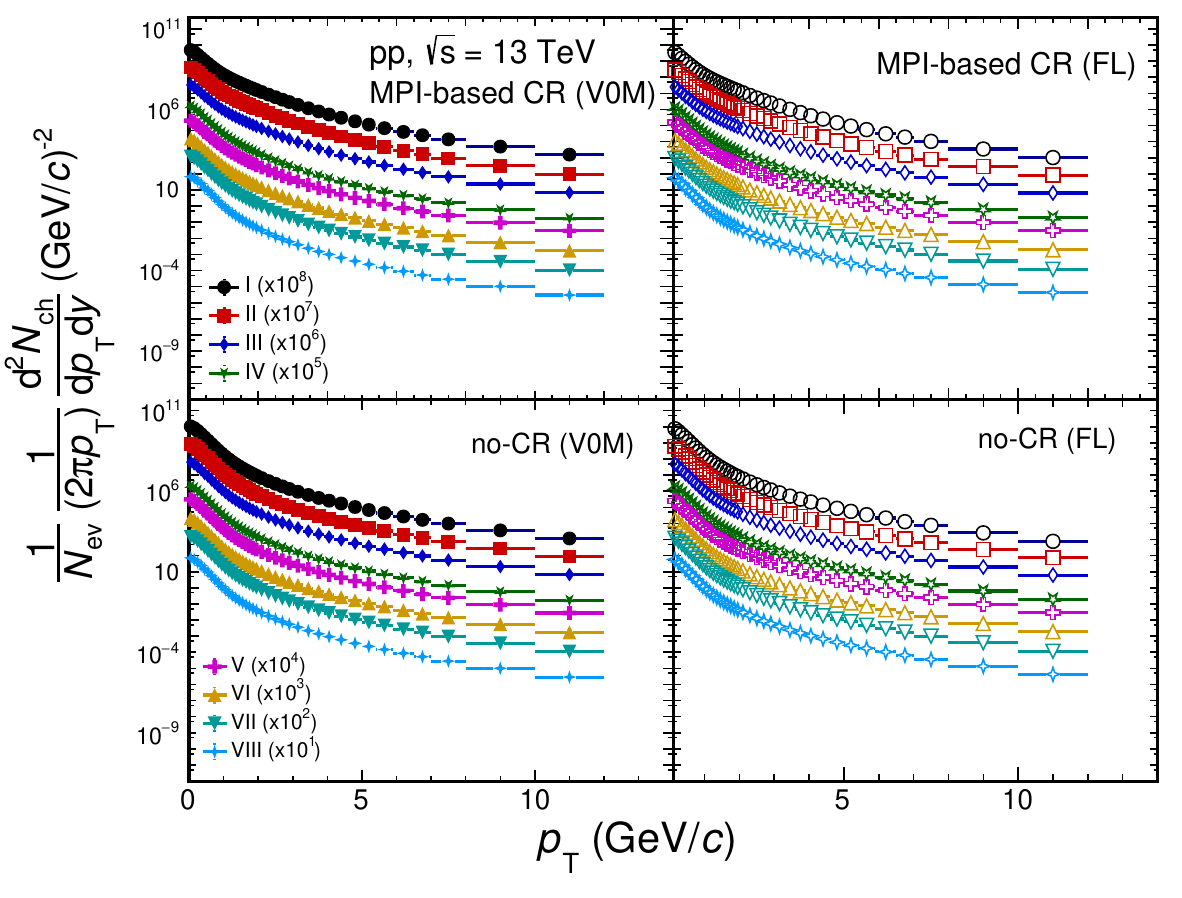}
    \caption{Transverse momentum distribution of charged particles in different multiplicity classes selected using the V0M and flattenicity (FL) estimators for CR and no-CR scenarios of \texttt{PYTHIA8} in $p+p$ collisions at $\sqrt{s} = 13$ TeV. The spectra are scaled by a factor $2^{n}$ for better visibility. Full and open markers represent the spectra for V0M and the flattenicity estimator respectively. }
    \label{fig:pTdist}
\end{figure*}
\vspace{-1pt}
The transverse momentum distribution ($p_{\rm T}$) of the charged particles produced in $p+p$ collisions at $\sqrt{s} = 13$ TeV is plotted in Fig.~\ref{fig:pTdist} for the two event classifiers. In Fig.~\ref{fig:pTdist}, the top and bottom panels represent the transverse momentum spectra of charged particles in multiplicity classes with and without CR, respectively. One can clearly see a multiplicity dependence of the spectral shapes obtained by using the V0M and flattenicity estimators, both with the studied variants (CR and no-CR) of the \texttt{PYTHIA8} model. As expected, the $p_{\rm T}$-spectra become harder as we move from low to high V0M multiplicity classes. This observation is in line with the ALICE experimental data \cite{Acharya_2019}. A similar pattern is also observed in the case of the flattenicity multiplicity estimator. \\~\\
\vspace{-1pt}
The nuclear modification factor ($R_{\rm AA}$ or $R_{\rm CP}$) in heavy-ion collisions is used to quantify the in-medium modification of the charged-particle transverse momentum spectrum \cite{2018}.
As already discussed in Section~\ref{intro}, the obscurity of determining the scaling factor in $p+p$ collisions motivates us to explore the other possibilities. One of the possibilities is to use the average number of charged pions, $\langle N_{\pi} \rangle$, as a measure of centrality in $A+A$ collisions, as reported in Refs. \cite{Scherer_1999, Dey_2017}. Authors in these references argued that the average number of participating nucleons, $\langle N_{\rm part} \rangle$, shows a nonlinear relationship with the volume of the participant zone, whereas $\langle N_{\pi} \rangle$ exhibits perfect participant scaling. In the current analysis, we therefore consider the average number of charged pions, $\langle N_{\pi} \rangle$, as the scaling factor. As already asserted in Section~\ref{intro}, the scaling factor also depends on the choice of event activity estimator. The V0M is the most common event classifier in $p+p$ collisions used by the ALICE Collaboration. It is worth mentioning that the V0M event classifier has an inherent bias towards hard processes, which makes it difficult to study the jet quenching effect in HM $p+p$ collisions. On the other hand, the new event classifier, flattenicity, has contributions from both soft and hard processes making it suitable to study transverse momentum distributions and hence the study of a possible jet quenching-like effect in terms of $\mathcal{P}_{\rm CP}$ in HM events in small systems ~\cite{Ortiz_2023}. \\~\\
\vspace{-1pt}
The quantity denoted by $\mathcal{P}_{\rm CP}$ for a small system such as $p + p$ collisions, which is analogous to the nuclear modification factor in $A + A$ collisions, is 
given by the following equation \cite{Dey:2022},

\begin{equation}
\mathcal{P}_{\rm CP} = \frac{\left[(d^{2}N_{\rm ch}/dp_{\rm T}dy)/\langle N_{\pi} \rangle \right]_{\rm HM}}{\left[(d^{2}N_{\rm ch}/dp_{\rm T}dy)/\langle N_{\pi} \rangle\right]_{\rm LM}}
 \label{Eq:Pcp}
\end{equation}
where the term $\langle N_{\pi} \rangle$ corresponds to the average number of charged pions in mid-rapidity. HM and LM in Eqn.~\ref{Eq:Pcp} correspond to the highest ($0-1\%$) and lowest ($50-100\%$) multiplicity classes, respectively. The $\mathcal{P}_{\rm CP}$ estimated using the two studied event classifiers for charged particles as a function of transverse momentum is shown in Fig.~\ref{fig:Pcp} using \texttt{PYTHIA8} with MPI-based CR (Monash).\\~\\
\vspace{-1pt}
It is clearly seen from  Fig.~\ref{fig:Pcp} that in the low-$p_{\rm T}$ region, the value of $\mathcal{P}_{\rm CP}$ for both the estimators are very close to each other and show a sharp increase
with $p_{\rm T}$, whereas beyond $p_{\rm T}=2$ GeV/c, a completely different trend is observed.
On the one hand, the $\mathcal{P}_{\rm CP}$ for the V0M event classifier exhibits a continuous rise with increasing $p_{\rm T}$ and saturates thereafter within the statistical uncertainties. On the other hand, the value of $\mathcal{P}_{\rm CP}$ estimated from flattenicity shows a bump-like structure i.e. the $\mathcal{P}_{\rm CP}$ values first increase and reach the maximum in the intermediate $p_{\rm T}$ ($2 - 4$ GeV/$c$) and then decrease at higher $p_{\rm T}$. This bump-like structure observed for flattenicity is hidden in the case of the V0M classifier. \\~\\
\vspace{-1pt}
In order to understand the origin of the observed bump-like structure, a comparative study has been performed where $\mathcal{P}_{\rm CP}$ has been estimated using the no-CR scenario for the two event classifiers and plotted in Fig.~\ref{fig:Pcp}. It is observed from the figure that the no-CR scenario reveals a monotonic increase in the value of $\mathcal{P}_{\rm cp}$ with $p_{\rm T}$ for V0M, whereas it saturates in the higher $p_{\rm T}$ region for the flattenicity classifier. The bump-like structure observed in the intermediate $p_{\rm T}$ is, therefore, attributed to the color reconnection (CR) mechanism implemented in the \texttt{PYTHIA8} model. It is also seen from Fig.~\ref{fig:Pcp} that beyond $p_{\rm T}$ = 6 GeV/$c$, a slight increase  of $\mathcal{P}_{\rm CP}$  with $p_{\rm T}$ for V0M is observed for both CR and no-CR modes. This observation is attributed to the presence of harder processes whose effects are more prominent at higher $p_{\rm T}$. 

\section{Summary}\label{Summary}
This letter reports on the effects of event classifiers on measuring jet quenching-like phenomena in $p+p$ collisions at $\sqrt{s} = 13$ TeV using a quantity $\mathcal{P}_{\rm CP}$  analogous to a standard observable like the nuclear modification factor ($R_{\rm CP}$). A new multiplicity classifier known as flattenicity is employed in the present study to estimate $\mathcal{P}_{\rm CP}$ and the results are compared with the standard V0M classifier. The study is performed using various tunes of \texttt{PYTHIA8} model for $p+p$ collisions at $\sqrt{s}=13$ TeV. A higher value of $\langle dN_{ch}/d\eta \rangle$ has been observed for the V0M method for HM events in comparison to the flattenicity technique for both CR and no-CR scenarios. The transverse momentum distribution shows the usual multiplicity-dependent evolution of spectral shapes (i.e., hardening of the spectra with increasing multiplicity) for the flattenicity estimator and is consistent with the trend observed for the V0M estimator. A bump-like structure for $\mathcal{P}_{\rm CP}$ estimated from flattenicity has been observed in the intermediate $p_{\rm T}$ region. No such structure is found when estimating  $\mathcal{P}_{\rm CP}$ using the V0M classifier. It is worth mentioning that such a bump-like structure has been reported in the Refs. \cite{Dey:2022, Ortiz_2020} where the multiplicity is selected in terms of the number of multi-parton interactions ($N_{\rm mpi}$). 
With the help of a new event classifier, known as flattenicity, we have therefore demonstrated that MPI (an initial state phenomenon) and CR (a final state effect) are responsible for the observed bump-like structure in $\mathcal{P}_{\rm CP}$. It is important to highlight here that the flattenicity classifier also selects events with contribution from lower transverse momentum parton-parton scatterings and results in a softer $p_{\rm T}$ spectrum in HM $p+p$ collisions than the V0M estimator.  \\~\\
\vspace{-1pt}
The present study reveals that the value of $\mathcal{P}_{\rm CP}$ is higher than unity for both the event classifiers in $p+p$ collisions at $\sqrt{s} = 13$ TeV using \texttt{PYTHIA8} generator. This result is consistent with the fact that medium formation is not implemented in \texttt{PYTHIA8} model. Therefore, it would be interesting to compare our current findings to the experimental data.


\section*{Acknowledgements}
The authors would like to thank Prof. D. K. Srivastava for careful reading of the manuscript and for valuable comments on it. Authors also thank Prof. Sukalyan Chattopadhyay and Dr. Mohd. Danish Azmi for their constructive suggestions. Authors would like to acknowledge the help received from Dipankar Basak for generating a part of the MC data. 

\bibliographystyle{unsrt} 
\bibliographystyle{model1-num-names}

\begin{thebibliography}{44}%
\makeatletter
\providecommand \@ifxundefined [1]{%
 \@ifx{#1\undefined}
}%
\providecommand \@ifnum [1]{%
 \ifnum #1\expandafter \@firstoftwo
 \else \expandafter \@secondoftwo
 \fi
}%
\providecommand \@ifx [1]{%
 \ifx #1\expandafter \@firstoftwo
 \else \expandafter \@secondoftwo
 \fi
}%
\providecommand \natexlab [1]{#1}%
\providecommand \enquote  [1]{``#1''}%
\providecommand \bibnamefont  [1]{#1}%
\providecommand \bibfnamefont [1]{#1}%
\providecommand \citenamefont [1]{#1}%
\providecommand \href@noop [0]{\@secondoftwo}%
\providecommand \href [0]{\begingroup \@sanitize@url \@href}%
\providecommand \@href[1]{\@@startlink{#1}\@@href}%
\providecommand \@@href[1]{\endgroup#1\@@endlink}%
\providecommand \@sanitize@url [0]{\catcode `\\12\catcode `\$12\catcode
  `\&12\catcode `\#12\catcode `\^12\catcode `\_12\catcode `\%12\relax}%
\providecommand \@@startlink[1]{}%
\providecommand \@@endlink[0]{}%
\providecommand \url  [0]{\begingroup\@sanitize@url \@url }%
\providecommand \@url [1]{\endgroup\@href {#1}{\urlprefix }}%
\providecommand \urlprefix  [0]{URL }%
\providecommand \Eprint [0]{\href }%
\providecommand \doibase [0]{http://dx.doi.org/}%
\providecommand \selectlanguage [0]{\@gobble}%
\providecommand \bibinfo  [0]{\@secondoftwo}%
\providecommand \bibfield  [0]{\@secondoftwo}%
\providecommand \translation [1]{[#1]}%
\providecommand \BibitemOpen [0]{}%
\providecommand \bibitemStop [0]{}%
\providecommand \bibitemNoStop [0]{.\EOS\space}%
\providecommand \EOS [0]{\spacefactor3000\relax}%
\providecommand \BibitemShut  [1]{\csname bibitem#1\endcsname}%
\let\auto@bib@innerbib\@empty
\bibitem [{\citenamefont {Adamczyk}\ \emph {et~al.}(2013)\citenamefont
  {Adamczyk} \emph {et~al.}}]{Adamczyk_2013}%
  \BibitemOpen
  \bibfield  {author} {\bibinfo {author} {\bibfnamefont {L.}~\bibnamefont
  {Adamczyk}} \emph {et~al.} (\bibinfo {collaboration} {CMS Collaboration}),\
  }\href {\doibase 10.1103/physrevlett.110.142301} {\bibfield  {journal}
  {\bibinfo  {journal} {Physical Review Letters}\ }\textbf {\bibinfo {volume}
  {110}},\ \bibinfo {pages} {142301} (\bibinfo {year} {2013})}\BibitemShut
  {NoStop}%
\bibitem [{\citenamefont {Drees}(2002)}]{Drees_2002}%
  \BibitemOpen
  \bibfield  {author} {\bibinfo {author} {\bibfnamefont {A.}~\bibnamefont
  {Drees}},\ }\href {\doibase 10.1016/s0375-9474(01)01380-x} {\bibfield
  {journal} {\bibinfo  {journal} {Nuclear Physics A}\ }\textbf {\bibinfo
  {volume} {698}},\ \bibinfo {pages} {331} (\bibinfo {year}
  {2002})}\BibitemShut {NoStop}%
\bibitem [{\citenamefont {Adam}\ \emph {et~al.}(2017)\citenamefont {Adam} \emph
  {et~al.}}]{nature:2017}%
  \BibitemOpen
  \bibfield  {author} {\bibinfo {author} {\bibfnamefont {J.}~\bibnamefont
  {Adam}} \emph {et~al.} (\bibinfo {collaboration} {ALICE Collaboration}),\
  }\href {\doibase 10.1038/nphys4111} {\bibfield  {journal} {\bibinfo
  {journal} {Nature Physics}\ }\textbf {\bibinfo {volume} {13}},\ \bibinfo
  {pages} {535} (\bibinfo {year} {2017})}\BibitemShut {NoStop}%
\bibitem [{\citenamefont {Khachatryan}\ \emph {et~al.}(2010)\citenamefont
  {Khachatryan} \emph {et~al.}}]{Khachatryan_2010}%
  \BibitemOpen
  \bibfield  {author} {\bibinfo {author} {\bibfnamefont {V.}~\bibnamefont
  {Khachatryan}} \emph {et~al.} (\bibinfo {collaboration} {CMS}),\ }\href
  {\doibase 10.1007/JHEP09(2010)091} {\bibfield  {journal} {\bibinfo  {journal}
  {JHEP}\ }\textbf {\bibinfo {volume} {09}},\ \bibinfo {pages} {091} (\bibinfo
  {year} {2010})},\ \Eprint {http://arxiv.org/abs/1009.4122} {arXiv:1009.4122
  [hep-ex]} \BibitemShut {NoStop}%
\bibitem [{\citenamefont {Khachatryan}\ \emph {et~al.}(2016)\citenamefont
  {Khachatryan} \emph {et~al.}}]{PhysRevLett.Khachatryan:2016}%
  \BibitemOpen
  \bibfield  {author} {\bibinfo {author} {\bibfnamefont {V.}~\bibnamefont
  {Khachatryan}} \emph {et~al.} (\bibinfo {collaboration} {CMS
  Collaboration}),\ }\href {\doibase 10.1103/PhysRevLett.116.172302} {\bibfield
   {journal} {\bibinfo  {journal} {Phys. Rev. Lett.}\ }\textbf {\bibinfo
  {volume} {116}},\ \bibinfo {pages} {172302} (\bibinfo {year}
  {2016})}\BibitemShut {NoStop}%
\bibitem [{\citenamefont {Velicanu}\ and\ \citenamefont {(forthe
  CMS~collaboration)}(2011)}]{Velicanu:2011}%
  \BibitemOpen
  \bibfield  {author} {\bibinfo {author} {\bibfnamefont {D.}~\bibnamefont
  {Velicanu}}\ and\ \bibinfo {author} {\bibnamefont {(forthe
  CMS~collaboration)}},\ }\href {\doibase 10.1088/0954-3899/38/12/124051}
  {\bibfield  {journal} {\bibinfo  {journal} {Journal of Physics G: Nuclear and
  Particle Physics}\ }\textbf {\bibinfo {volume} {38}},\ \bibinfo {pages}
  {124051} (\bibinfo {year} {2011})}\BibitemShut {NoStop}%
\bibitem [{\citenamefont {Aad}\ \emph {et~al.}(2016)\citenamefont {Aad} \emph
  {et~al.}}]{PhysRevLett.Aad:2016}%
  \BibitemOpen
  \bibfield  {author} {\bibinfo {author} {\bibfnamefont {G.}~\bibnamefont
  {Aad}} \emph {et~al.} (\bibinfo {collaboration} {ATLAS Collaboration}),\
  }\href {\doibase 10.1103/PhysRevLett.116.172301} {\bibfield  {journal}
  {\bibinfo  {journal} {Phys. Rev. Lett.}\ }\textbf {\bibinfo {volume} {116}},\
  \bibinfo {pages} {172301} (\bibinfo {year} {2016})}\BibitemShut {NoStop}%
\bibitem [{\citenamefont {Miller}\ \emph {et~al.}(2007)\citenamefont {Miller}
  \emph {et~al.}}]{Miller:2017}%
  \BibitemOpen
  \bibfield  {author} {\bibinfo {author} {\bibfnamefont {M.~L.}\ \bibnamefont
  {Miller}} \emph {et~al.},\ }\href {\doibase
  10.1146/annurev.nucl.57.090506.123020} {\bibfield  {journal} {\bibinfo
  {journal} {Annual Review of Nuclear and Particle Science}\ }\textbf {\bibinfo
  {volume} {57}},\ \bibinfo {pages} {205} (\bibinfo {year} {2007})}\BibitemShut
  {NoStop}%
\bibitem [{\citenamefont {Acharya}\ \emph
  {et~al.}(2019{\natexlab{a}})\citenamefont {Acharya} \emph
  {et~al.}}]{PhysRevC.99.024906:auto}%
  \BibitemOpen
  \bibfield  {author} {\bibinfo {author} {\bibfnamefont {S.}~\bibnamefont
  {Acharya}} \emph {et~al.} (\bibinfo {collaboration} {ALICE Collaboration}),\
  }\href {\doibase 10.1103/PhysRevC.99.024906} {\bibfield  {journal} {\bibinfo
  {journal} {Phys. Rev. C}\ }\textbf {\bibinfo {volume} {99}},\ \bibinfo
  {pages} {024906} (\bibinfo {year} {2019}{\natexlab{a}})}\BibitemShut
  {NoStop}%
\bibitem [{\citenamefont {Abelev}\ \emph
  {et~al.}(2012{\natexlab{a}})\citenamefont {Abelev} \emph
  {et~al.}}]{Abelev_2012}%
  \BibitemOpen
  \bibfield  {author} {\bibinfo {author} {\bibfnamefont {B.}~\bibnamefont
  {Abelev}} \emph {et~al.} (\bibinfo {collaboration} {ALICE Collaboration}),\
  }\href {\doibase 10.1140/epjc/s10052-012-2124-9} {\bibfield  {journal}
  {\bibinfo  {journal} {The European Physical Journal C}\ }\textbf {\bibinfo
  {volume} {72}},\ \bibinfo {pages} {2124} (\bibinfo {year}
  {2012}{\natexlab{a}})}\BibitemShut {NoStop}%
\bibitem [{\citenamefont {Acharya}\ \emph
  {et~al.}(2019{\natexlab{b}})\citenamefont {Acharya} \emph
  {et~al.}}]{Acharya_2019}%
  \BibitemOpen
  \bibfield  {author} {\bibinfo {author} {\bibfnamefont {S.}~\bibnamefont
  {Acharya}} \emph {et~al.} (\bibinfo {collaboration} {ALICE Collaboration}),\
  }\href {\doibase 10.1140/epjc/s10052-019-7350-y} {\bibfield  {journal}
  {\bibinfo  {journal} {The European Physical Journal C}\ }\textbf {\bibinfo
  {volume} {79}},\ \bibinfo {pages} {857} (\bibinfo {year}
  {2019}{\natexlab{b}})}\BibitemShut {NoStop}%
\bibitem [{\citenamefont {Ortiz}\ \emph {et~al.}(2015)\citenamefont {Ortiz},
  \citenamefont {Cuautle},\ and\ \citenamefont {Pai$\acute{c}$}}]{Ortiz_2015}%
  \BibitemOpen
  \bibfield  {author} {\bibinfo {author} {\bibfnamefont {A.}~\bibnamefont
  {Ortiz}}, \bibinfo {author} {\bibfnamefont {E.}~\bibnamefont {Cuautle}}, \
  and\ \bibinfo {author} {\bibfnamefont {G.}~\bibnamefont {Pai$\acute{c}$}},\
  }\href {\doibase 10.1016/j.nuclphysa.2015.05.010} {\bibfield  {journal}
  {\bibinfo  {journal} {Nuclear Physics A}\ }\textbf {\bibinfo {volume}
  {941}},\ \bibinfo {pages} {78} (\bibinfo {year} {2015})}\BibitemShut
  {NoStop}%
\bibitem [{\citenamefont {Martin}\ \emph {et~al.}(2016)\citenamefont {Martin},
  \citenamefont {Skands},\ and\ \citenamefont {Farrington}}]{Martin_2016}%
  \BibitemOpen
  \bibfield  {author} {\bibinfo {author} {\bibfnamefont {T.}~\bibnamefont
  {Martin}}, \bibinfo {author} {\bibfnamefont {P.}~\bibnamefont {Skands}}, \
  and\ \bibinfo {author} {\bibfnamefont {S.}~\bibnamefont {Farrington}},\
  }\href {\doibase 10.1140/epjc/s10052-016-4135-4} {\bibfield  {journal}
  {\bibinfo  {journal} {The European Physical Journal C}\ }\textbf {\bibinfo
  {volume} {76}},\ \bibinfo {pages} {299} (\bibinfo {year} {2016})}\BibitemShut
  {NoStop}%
\bibitem [{\citenamefont {Ortiz}\ and\ \citenamefont
  {Palomo}(2017)}]{Ortiz_2017}%
  \BibitemOpen
  \bibfield  {author} {\bibinfo {author} {\bibfnamefont {A.}~\bibnamefont
  {Ortiz}}\ and\ \bibinfo {author} {\bibfnamefont {L.~V.}\ \bibnamefont
  {Palomo}},\ }\href {\doibase 10.1103/physrevd.96.114019} {\bibfield
  {journal} {\bibinfo  {journal} {Physical Review D}\ }\textbf {\bibinfo
  {volume} {96}},\ \bibinfo {pages} {114019} (\bibinfo {year}
  {2017})}\BibitemShut {NoStop}%
\bibitem [{\citenamefont {Ortiz}\ \emph {et~al.}(2023)\citenamefont {Ortiz},
  \citenamefont {Khuntia}, \citenamefont {Rueda}, \citenamefont {Tripathy},
  \citenamefont {Benc$\acute{e}$di}, \citenamefont {Prasad},\ and\
  \citenamefont {Fan}}]{Ortiz_2023}%
  \BibitemOpen
  \bibfield  {author} {\bibinfo {author} {\bibfnamefont {A.}~\bibnamefont
  {Ortiz}}, \bibinfo {author} {\bibfnamefont {A.}~\bibnamefont {Khuntia}},
  \bibinfo {author} {\bibfnamefont {O.~V.}\ \bibnamefont {Rueda}}, \bibinfo
  {author} {\bibfnamefont {S.}~\bibnamefont {Tripathy}}, \bibinfo {author}
  {\bibfnamefont {G.}~\bibnamefont {Benc$\acute{e}$di}}, \bibinfo {author}
  {\bibfnamefont {S.}~\bibnamefont {Prasad}}, \ and\ \bibinfo {author}
  {\bibfnamefont {F.}~\bibnamefont {Fan}},\ }\href {\doibase
  10.1103/physrevd.107.076012} {\bibfield  {journal} {\bibinfo  {journal}
  {Physical Review D}\ }\textbf {\bibinfo {volume} {107}},\ \bibinfo {pages}
  {076012} (\bibinfo {year} {2023})}\BibitemShut {NoStop}%
\bibitem [{\citenamefont {Ortiz}\ and\ \citenamefont
  {Paic}(2022)}]{ortiz2022look}%
  \BibitemOpen
  \bibfield  {author} {\bibinfo {author} {\bibfnamefont {A.}~\bibnamefont
  {Ortiz}}\ and\ \bibinfo {author} {\bibfnamefont {G.}~\bibnamefont {Paic}},\
  }\href {\doibase 10.31349/SuplRevMexFis.3.040911} {\bibfield  {journal}
  {\bibinfo  {journal} {Suplemento de la Revista Mexicana de Física}\ }\textbf
  {\bibinfo {volume} {3}},\ \bibinfo {pages} {040911} (\bibinfo {year}
  {2022})},\ \Eprint {http://arxiv.org/abs/2204.13733} {arXiv:2204.13733
  [hep-ph]} \BibitemShut {NoStop}%
\bibitem [{\citenamefont {Dey}\ \emph {et~al.}(2022)\citenamefont {Dey},
  \citenamefont {Mishra},\ and\ \citenamefont {Tripathy}}]{Dey:2022}%
  \BibitemOpen
  \bibfield  {author} {\bibinfo {author} {\bibfnamefont {K.}~\bibnamefont
  {Dey}}, \bibinfo {author} {\bibfnamefont {A.~N.}\ \bibnamefont {Mishra}}, \
  and\ \bibinfo {author} {\bibfnamefont {S.~K.}\ \bibnamefont {Tripathy}},\
  }\href {\doibase 10.1209/0295-5075/ac535b} {\bibfield  {journal} {\bibinfo
  {journal} {Europhysics Letters}\ }\textbf {\bibinfo {volume} {136}},\
  \bibinfo {pages} {62001} (\bibinfo {year} {2022})}\BibitemShut {NoStop}%
\bibitem [{\citenamefont {Sj$\ddot{o}$strand}\ \emph
  {et~al.}(2006)\citenamefont {Sj$\ddot{o}$strand}, \citenamefont {Mrenna},\
  and\ \citenamefont {Skands}}]{Sj_strand_2006}%
  \BibitemOpen
  \bibfield  {author} {\bibinfo {author} {\bibfnamefont {T.}~\bibnamefont
  {Sj$\ddot{o}$strand}}, \bibinfo {author} {\bibfnamefont {S.}~\bibnamefont
  {Mrenna}}, \ and\ \bibinfo {author} {\bibfnamefont {P.}~\bibnamefont
  {Skands}},\ }\href {\doibase 10.1088/1126-6708/2006/05/026} {\bibfield
  {journal} {\bibinfo  {journal} {Journal of High Energy Physics}\ }\textbf
  {\bibinfo {volume} {2006}},\ \bibinfo {pages} {026} (\bibinfo {year}
  {2006})}\BibitemShut {NoStop}%
\bibitem [{\citenamefont {Sj$\ddot{o}$strand}\ \emph
  {et~al.}(2008)\citenamefont {Sj$\ddot{o}$strand}, \citenamefont {Mrenna},\
  and\ \citenamefont {Skands}}]{Sj_strand_2008}%
  \BibitemOpen
  \bibfield  {author} {\bibinfo {author} {\bibfnamefont {T.}~\bibnamefont
  {Sj$\ddot{o}$strand}}, \bibinfo {author} {\bibfnamefont {S.}~\bibnamefont
  {Mrenna}}, \ and\ \bibinfo {author} {\bibfnamefont {P.}~\bibnamefont
  {Skands}},\ }\href {\doibase 10.1016/j.cpc.2008.01.036} {\bibfield  {journal}
  {\bibinfo  {journal} {Computer Physics Communications}\ }\textbf {\bibinfo
  {volume} {178}},\ \bibinfo {pages} {852} (\bibinfo {year}
  {2008})}\BibitemShut {NoStop}%
\bibitem [{\citenamefont {Sj$\ddot{o}$strand}\ \emph
  {et~al.}(2015)\citenamefont {Sj$\ddot{o}$strand}, \citenamefont {Ask},
  \citenamefont {Christiansen}, \citenamefont {Corke}, \citenamefont {Desai},
  \citenamefont {Ilten}, \citenamefont {Mrenna}, \citenamefont {Prestel},
  \citenamefont {Rasmussen},\ and\ \citenamefont {Skands}}]{Sj_strand_2015}%
  \BibitemOpen
  \bibfield  {author} {\bibinfo {author} {\bibfnamefont {T.}~\bibnamefont
  {Sj$\ddot{o}$strand}}, \bibinfo {author} {\bibfnamefont {S.}~\bibnamefont
  {Ask}}, \bibinfo {author} {\bibfnamefont {J.~R.}\ \bibnamefont
  {Christiansen}}, \bibinfo {author} {\bibfnamefont {R.}~\bibnamefont {Corke}},
  \bibinfo {author} {\bibfnamefont {N.}~\bibnamefont {Desai}}, \bibinfo
  {author} {\bibfnamefont {P.}~\bibnamefont {Ilten}}, \bibinfo {author}
  {\bibfnamefont {S.}~\bibnamefont {Mrenna}}, \bibinfo {author} {\bibfnamefont
  {S.}~\bibnamefont {Prestel}}, \bibinfo {author} {\bibfnamefont {C.~O.}\
  \bibnamefont {Rasmussen}}, \ and\ \bibinfo {author} {\bibfnamefont {P.~Z.}\
  \bibnamefont {Skands}},\ }\href {\doibase 10.1016/j.cpc.2015.01.024}
  {\bibfield  {journal} {\bibinfo  {journal} {Computer Physics Communications}\
  }\textbf {\bibinfo {volume} {191}},\ \bibinfo {pages} {159} (\bibinfo {year}
  {2015})}\BibitemShut {NoStop}%
\bibitem [{\citenamefont {Bierlich}\ \emph {et~al.}(2022)\citenamefont
  {Bierlich}, \citenamefont {Chakraborty}, \citenamefont {Desai}, \citenamefont
  {Gellersen}, \citenamefont {Helenius}, \citenamefont {Ilten}, \citenamefont
  {Lönnblad}, \citenamefont {Mrenna}, \citenamefont {Prestel}, \citenamefont
  {Preuss}, \citenamefont {Sj$\ddot{o}$strand}, \citenamefont {Skands},
  \citenamefont {Utheim},\ and\ \citenamefont
  {Verheyen}}]{10.21468/SciPostPhysCodeb.8}%
  \BibitemOpen
  \bibfield  {author} {\bibinfo {author} {\bibfnamefont {C.}~\bibnamefont
  {Bierlich}}, \bibinfo {author} {\bibfnamefont {S.}~\bibnamefont
  {Chakraborty}}, \bibinfo {author} {\bibfnamefont {N.}~\bibnamefont {Desai}},
  \bibinfo {author} {\bibfnamefont {L.}~\bibnamefont {Gellersen}}, \bibinfo
  {author} {\bibfnamefont {I.}~\bibnamefont {Helenius}}, \bibinfo {author}
  {\bibfnamefont {P.}~\bibnamefont {Ilten}}, \bibinfo {author} {\bibfnamefont
  {L.}~\bibnamefont {Lönnblad}}, \bibinfo {author} {\bibfnamefont
  {S.}~\bibnamefont {Mrenna}}, \bibinfo {author} {\bibfnamefont
  {S.}~\bibnamefont {Prestel}}, \bibinfo {author} {\bibfnamefont {C.~T.}\
  \bibnamefont {Preuss}}, \bibinfo {author} {\bibfnamefont {T.}~\bibnamefont
  {Sj$\ddot{o}$strand}}, \bibinfo {author} {\bibfnamefont {P.}~\bibnamefont
  {Skands}}, \bibinfo {author} {\bibfnamefont {M.}~\bibnamefont {Utheim}}, \
  and\ \bibinfo {author} {\bibfnamefont {R.}~\bibnamefont {Verheyen}},\ }\href
  {\doibase 10.21468/SciPostPhysCodeb.8} {\bibfield  {journal} {\bibinfo
  {journal} {SciPost Physics Codebases}\ ,\ \bibinfo {pages} {8}} (\bibinfo
  {year} {2022})}\BibitemShut {NoStop}%
\bibitem [{\citenamefont {Abazov}\ \emph {et~al.}(2010)\citenamefont {Abazov}
  \emph {et~al.}}]{Abazov_2010}%
  \BibitemOpen
  \bibfield  {author} {\bibinfo {author} {\bibfnamefont {V.~M.}\ \bibnamefont
  {Abazov}} \emph {et~al.} (\bibinfo {collaboration} {D0 Collaboration}),\
  }\href {\doibase 10.1103/physrevd.81.052012} {\bibfield  {journal} {\bibinfo
  {journal} {Physical Review D}\ }\textbf {\bibinfo {volume} {81}},\ \bibinfo
  {pages} {052012} (\bibinfo {year} {2010})}\BibitemShut {NoStop}%
\bibitem [{\citenamefont {Chekanov}\ \emph {et~al.}(2008)\citenamefont
  {Chekanov} \emph {et~al.}}]{Chekanov_2008}%
  \BibitemOpen
  \bibfield  {author} {\bibinfo {author} {\bibfnamefont {S.}~\bibnamefont
  {Chekanov}} \emph {et~al.} (\bibinfo {collaboration} {ZEUS Collaboration}),\
  }\href {\doibase 10.1016/j.nuclphysb.2007.08.021} {\bibfield  {journal}
  {\bibinfo  {journal} {Nuclear Physics B}\ }\textbf {\bibinfo {volume}
  {792}},\ \bibinfo {pages} {1} (\bibinfo {year} {2008})}\BibitemShut {NoStop}%
\bibitem [{\citenamefont {Adam}\ \emph {et~al.}(2016)\citenamefont {Adam} \emph
  {et~al.}}]{Adam_2016}%
  \BibitemOpen
  \bibfield  {author} {\bibinfo {author} {\bibfnamefont {J.}~\bibnamefont
  {Adam}} \emph {et~al.} (\bibinfo {collaboration} {ALICE Collaboration}),\
  }\href {\doibase 10.1016/j.physletb.2015.12.030} {\bibfield  {journal}
  {\bibinfo  {journal} {Physics Letters B}\ }\textbf {\bibinfo {volume}
  {753}},\ \bibinfo {pages} {319} (\bibinfo {year} {2016})}\BibitemShut
  {NoStop}%
\bibitem [{\citenamefont {Abelev}\ \emph
  {et~al.}(2012{\natexlab{b}})\citenamefont {Abelev} \emph
  {et~al.}}]{Abelev_2012_strange}%
  \BibitemOpen
  \bibfield  {author} {\bibinfo {author} {\bibfnamefont {B.}~\bibnamefont
  {Abelev}} \emph {et~al.} (\bibinfo {collaboration} {ALICE Collaboration}),\
  }\href {\doibase 10.1140/epjc/s10052-012-2183-y} {\bibfield  {journal}
  {\bibinfo  {journal} {The European Physical Journal C}\ }\textbf {\bibinfo
  {volume} {72}},\ \bibinfo {pages} {183} (\bibinfo {year}
  {2012}{\natexlab{b}})}\BibitemShut {NoStop}%
\bibitem [{\citenamefont {Acharya}\ \emph
  {et~al.}(2022{\natexlab{a}})\citenamefont {Acharya} \emph
  {et~al.}}]{Acharya_2022}%
  \BibitemOpen
  \bibfield  {author} {\bibinfo {author} {\bibfnamefont {S.}~\bibnamefont
  {Acharya}} \emph {et~al.} (\bibinfo {collaboration} {ALICE Collaboration}),\
  }\href {\doibase 10.1140/epjc/s10052-022-10405-x} {\bibfield  {journal}
  {\bibinfo  {journal} {The European Physical Journal C}\ }\textbf {\bibinfo
  {volume} {82}},\ \bibinfo {pages} {514} (\bibinfo {year}
  {2022}{\natexlab{a}})}\BibitemShut {NoStop}%
\bibitem [{\citenamefont {Argyropoulos}\ and\ \citenamefont
  {Sj$\ddot{o}$strand}(2014)}]{Argyropoulos_2014}%
  \BibitemOpen
  \bibfield  {author} {\bibinfo {author} {\bibfnamefont {S.}~\bibnamefont
  {Argyropoulos}}\ and\ \bibinfo {author} {\bibfnamefont {T.}~\bibnamefont
  {Sj$\ddot{o}$strand}},\ }\href {\doibase 10.1007/jhep11(2014)043} {\bibfield
  {journal} {\bibinfo  {journal} {Journal of High Energy Physics}\ }\textbf
  {\bibinfo {volume} {2014}} (\bibinfo {year} {2014}),\
  10.1007/jhep11(2014)043}\BibitemShut {NoStop}%
\bibitem [{\citenamefont {Gustafson}(2009)}]{gustafson2009}%
  \BibitemOpen
  \bibfield  {author} {\bibinfo {author} {\bibfnamefont {G.}~\bibnamefont
  {Gustafson}},\ }\href
  {https://www.actaphys.uj.edu.pl/fulltext?series=Reg&vol=40&page=1981}
  {\bibfield  {journal} {\bibinfo  {journal} {Acta Physica Polonica B}\
  }\textbf {\bibinfo {volume} {40}},\ \bibinfo {pages} {1981 } (\bibinfo {year}
  {2009})}\BibitemShut {NoStop}%
\bibitem [{\citenamefont {Skands}\ \emph {et~al.}(2014)\citenamefont {Skands},
  \citenamefont {Carrazza},\ and\ \citenamefont {Rojo}}]{Skands_2014}%
  \BibitemOpen
  \bibfield  {author} {\bibinfo {author} {\bibfnamefont {P.}~\bibnamefont
  {Skands}}, \bibinfo {author} {\bibfnamefont {S.}~\bibnamefont {Carrazza}}, \
  and\ \bibinfo {author} {\bibfnamefont {J.}~\bibnamefont {Rojo}},\ }\href
  {\doibase 10.1140/epjc/s10052-014-3024-y} {\bibfield  {journal} {\bibinfo
  {journal} {The European Physical Journal C}\ }\textbf {\bibinfo {volume}
  {74}},\ \bibinfo {pages} {3024} (\bibinfo {year} {2014})}\BibitemShut
  {NoStop}%
\bibitem [{\citenamefont {Abelev}\ \emph {et~al.}(2013)\citenamefont {Abelev}
  \emph {et~al.}}]{Abelev_2013}%
  \BibitemOpen
  \bibfield  {author} {\bibinfo {author} {\bibfnamefont {B.}~\bibnamefont
  {Abelev}} \emph {et~al.} (\bibinfo {collaboration} {ALICE Collaboration}),\
  }\href {\doibase 10.1016/j.physletb.2013.10.054} {\bibfield  {journal}
  {\bibinfo  {journal} {Physics Letters B}\ }\textbf {\bibinfo {volume}
  {727}},\ \bibinfo {pages} {371} (\bibinfo {year} {2013})}\BibitemShut
  {NoStop}%
\bibitem [{\citenamefont {Abelev}\ \emph
  {et~al.}(2012{\natexlab{c}})\citenamefont {Abelev} \emph
  {et~al.}}]{Abelev_2012_Jpsi}%
  \BibitemOpen
  \bibfield  {author} {\bibinfo {author} {\bibfnamefont {B.}~\bibnamefont
  {Abelev}} \emph {et~al.} (\bibinfo {collaboration} {ALICE Collaboration}),\
  }\href {\doibase 10.1016/j.physletb.2012.04.052} {\bibfield  {journal}
  {\bibinfo  {journal} {Physics Letters B}\ }\textbf {\bibinfo {volume}
  {712}},\ \bibinfo {pages} {165} (\bibinfo {year}
  {2012}{\natexlab{c}})}\BibitemShut {NoStop}%
\bibitem [{\citenamefont {Thakur}\ \emph {et~al.}(2018)\citenamefont {Thakur},
  \citenamefont {De}, \citenamefont {Sahoo},\ and\ \citenamefont
  {Dansana}}]{Thakur_2018}%
  \BibitemOpen
  \bibfield  {author} {\bibinfo {author} {\bibfnamefont {D.}~\bibnamefont
  {Thakur}}, \bibinfo {author} {\bibfnamefont {S.}~\bibnamefont {De}}, \bibinfo
  {author} {\bibfnamefont {R.}~\bibnamefont {Sahoo}}, \ and\ \bibinfo {author}
  {\bibfnamefont {S.}~\bibnamefont {Dansana}},\ }\href {\doibase
  10.1103/physrevd.97.094002} {\bibfield  {journal} {\bibinfo  {journal}
  {Physical Review D}\ }\textbf {\bibinfo {volume} {97}},\ \bibinfo {pages}
  {094002} (\bibinfo {year} {2018})}\BibitemShut {NoStop}%
\bibitem [{\citenamefont {Ortiz~Velasquez}\ \emph {et~al.}(2013)\citenamefont
  {Ortiz~Velasquez}, \citenamefont {Christiansen}, \citenamefont
  {Cuautle~Flores}, \citenamefont {Maldonado~Cervantes},\ and\ \citenamefont
  {Pai\ifmmode~\acute{c}\else \'{c}\fi{}}}]{Ortiz_Velasquez_2013}%
  \BibitemOpen
  \bibfield  {author} {\bibinfo {author} {\bibfnamefont {A.}~\bibnamefont
  {Ortiz~Velasquez}}, \bibinfo {author} {\bibfnamefont {P.}~\bibnamefont
  {Christiansen}}, \bibinfo {author} {\bibfnamefont {E.}~\bibnamefont
  {Cuautle~Flores}}, \bibinfo {author} {\bibfnamefont {I.~A.}\ \bibnamefont
  {Maldonado~Cervantes}}, \ and\ \bibinfo {author} {\bibfnamefont
  {G.}~\bibnamefont {Pai\ifmmode~\acute{c}\else \'{c}\fi{}}},\ }\href {\doibase
  10.1103/PhysRevLett.111.042001} {\bibfield  {journal} {\bibinfo  {journal}
  {Phys. Rev. Lett.}\ }\textbf {\bibinfo {volume} {111}},\ \bibinfo {pages}
  {042001} (\bibinfo {year} {2013})}\BibitemShut {NoStop}%
\bibitem [{\citenamefont {Bierlich}\ and\ \citenamefont
  {Christiansen}(2015)}]{Bierlich_2015}%
  \BibitemOpen
  \bibfield  {author} {\bibinfo {author} {\bibfnamefont {C.}~\bibnamefont
  {Bierlich}}\ and\ \bibinfo {author} {\bibfnamefont {J.~R.}\ \bibnamefont
  {Christiansen}},\ }\href {\doibase 10.1103/physrevd.92.094010} {\bibfield
  {journal} {\bibinfo  {journal} {Physical Review D}\ }\textbf {\bibinfo
  {volume} {92}},\ \bibinfo {pages} {094010} (\bibinfo {year}
  {2015})}\BibitemShut {NoStop}%
\bibitem [{\citenamefont {Acharya}\ \emph {et~al.}(2021)\citenamefont {Acharya}
  \emph {et~al.}}]{Acharya_2021}%
  \BibitemOpen
  \bibfield  {author} {\bibinfo {author} {\bibfnamefont {S.}~\bibnamefont
  {Acharya}} \emph {et~al.} (\bibinfo {collaboration} {ALICE Collaboration}),\
  }\href {\doibase 10.1103/physrevc.104.054905} {\bibfield  {journal} {\bibinfo
   {journal} {Physical Review C}\ }\textbf {\bibinfo {volume} {104}},\ \bibinfo
  {pages} {054905} (\bibinfo {year} {2021})}\BibitemShut {NoStop}%
\bibitem [{\citenamefont {Acharya}\ \emph
  {et~al.}(2022{\natexlab{b}})\citenamefont {Acharya} \emph
  {et~al.}}]{Acharya_2022_baryon}%
  \BibitemOpen
  \bibfield  {author} {\bibinfo {author} {\bibfnamefont {S.}~\bibnamefont
  {Acharya}} \emph {et~al.} (\bibinfo {collaboration} {ALICE Collaboration}),\
  }\href {\doibase 10.1103/physrevlett.128.012001} {\bibfield  {journal}
  {\bibinfo  {journal} {Physical Review Letters}\ }\textbf {\bibinfo {volume}
  {128}},\ \bibinfo {pages} {012001} (\bibinfo {year}
  {2022}{\natexlab{b}})}\BibitemShut {NoStop}%
\bibitem [{\citenamefont {Abelev}\ \emph {et~al.}(2014)\citenamefont {Abelev}
  \emph {et~al.}}]{ALICE_2014}%
  \BibitemOpen
  \bibfield  {author} {\bibinfo {author} {\bibfnamefont {B.}~\bibnamefont
  {Abelev}} \emph {et~al.} (\bibinfo {collaboration} {ALICE Collaboration}),\
  }\href {\doibase 10.1142/s0217751x14300440} {\bibfield  {journal} {\bibinfo
  {journal} {International Journal of Modern Physics A}\ }\textbf {\bibinfo
  {volume} {29}},\ \bibinfo {pages} {1430044} (\bibinfo {year}
  {2014})}\BibitemShut {NoStop}%
\bibitem [{\citenamefont {Christiansen}\ and\ \citenamefont
  {Skands}(2015)}]{Christiansen_2015}%
  \BibitemOpen
  \bibfield  {author} {\bibinfo {author} {\bibfnamefont {J.~R.}\ \bibnamefont
  {Christiansen}}\ and\ \bibinfo {author} {\bibfnamefont {P.~Z.}\ \bibnamefont
  {Skands}},\ }\href {\doibase 10.1007/jhep08(2015)003} {\bibfield  {journal}
  {\bibinfo  {journal} {Journal of High Energy Physics}\ }\textbf {\bibinfo
  {volume} {2015}},\ \bibinfo {pages} {003} (\bibinfo {year}
  {2015})}\BibitemShut {NoStop}%
\bibitem [{\citenamefont {Adam}\ \emph {et~al.}(2018)\citenamefont {Adam} \emph
  {et~al.}}]{Adam_2018}%
  \BibitemOpen
  \bibfield  {author} {\bibinfo {author} {\bibfnamefont {J.}~\bibnamefont
  {Adam}} \emph {et~al.} (\bibinfo {collaboration} {STAR Collaboration}),\
  }\href {\doibase 10.1103/physrevd.98.112009} {\bibfield  {journal} {\bibinfo
  {journal} {Physical Review D}\ }\textbf {\bibinfo {volume} {98}},\ \bibinfo
  {pages} {112009} (\bibinfo {year} {2018})}\BibitemShut {NoStop}%
\bibitem [{\citenamefont {Khachatryan}\ \emph {et~al.}(2017)\citenamefont
  {Khachatryan} \emph {et~al.}}]{Khachatryan_2017}%
  \BibitemOpen
  \bibfield  {author} {\bibinfo {author} {\bibfnamefont {V.}~\bibnamefont
  {Khachatryan}} \emph {et~al.} (\bibinfo {collaboration} {CMS
  Collaboration}),\ }\href {\doibase 10.1016/j.physletb.2017.01.075} {\bibfield
   {journal} {\bibinfo  {journal} {Physics Letters B}\ }\textbf {\bibinfo
  {volume} {768}},\ \bibinfo {pages} {103} (\bibinfo {year}
  {2017})}\BibitemShut {NoStop}%
\bibitem [{\citenamefont {Acharya}\ \emph {et~al.}(2018)\citenamefont {Acharya}
  \emph {et~al.}}]{2018}%
  \BibitemOpen
  \bibfield  {author} {\bibinfo {author} {\bibfnamefont {S.}~\bibnamefont
  {Acharya}} \emph {et~al.} (\bibinfo {collaboration} {ALICE Collaboration}),\
  }\href {\doibase 10.1007/jhep11(2018)013} {\bibfield  {journal} {\bibinfo
  {journal} {Journal of High Energy Physics}\ }\textbf {\bibinfo {volume}
  {2018}},\ \bibinfo {pages} {013} (\bibinfo {year} {2018})}\BibitemShut
  {NoStop}%
\bibitem [{\citenamefont {Scherer}\ \emph {et~al.}(1999)\citenamefont {Scherer}
  \emph {et~al.}}]{Scherer_1999}%
  \BibitemOpen
  \bibfield  {author} {\bibinfo {author} {\bibfnamefont {S.}~\bibnamefont
  {Scherer}} \emph {et~al.},\ }\href {\doibase 10.1016/S0146-6410(99)00083-6}
  {\bibfield  {journal} {\bibinfo  {journal} {Progress in Particle and Nuclear
  Physics}\ }\textbf {\bibinfo {volume} {42}},\ \bibinfo {pages} {279}
  (\bibinfo {year} {1999})}\BibitemShut {NoStop}%
\bibitem [{\citenamefont {Dey}\ and\ \citenamefont
  {Bhattacharjee}(2017)}]{Dey_2017}%
  \BibitemOpen
  \bibfield  {author} {\bibinfo {author} {\bibfnamefont {K.}~\bibnamefont
  {Dey}}\ and\ \bibinfo {author} {\bibfnamefont {B.}~\bibnamefont
  {Bhattacharjee}},\ }\href {\doibase 10.1016/j.nuclphysa.2017.05.092}
  {\bibfield  {journal} {\bibinfo  {journal} {Nuclear Physics A}\ }\textbf
  {\bibinfo {volume} {965}},\ \bibinfo {pages} {74} (\bibinfo {year}
  {2017})}\BibitemShut {NoStop}%
\bibitem [{\citenamefont {Ortiz}\ \emph {et~al.}(2020)\citenamefont {Ortiz},
  \citenamefont {Paz}, \citenamefont {Romo}, \citenamefont {Tripathy},
  \citenamefont {Zepeda},\ and\ \citenamefont {Bautista}}]{Ortiz_2020}%
  \BibitemOpen
  \bibfield  {author} {\bibinfo {author} {\bibfnamefont {A.}~\bibnamefont
  {Ortiz}}, \bibinfo {author} {\bibfnamefont {A.}~\bibnamefont {Paz}}, \bibinfo
  {author} {\bibfnamefont {J.~D.}\ \bibnamefont {Romo}}, \bibinfo {author}
  {\bibfnamefont {S.}~\bibnamefont {Tripathy}}, \bibinfo {author}
  {\bibfnamefont {E.~A.}\ \bibnamefont {Zepeda}}, \ and\ \bibinfo {author}
  {\bibfnamefont {I.}~\bibnamefont {Bautista}},\ }\href {\doibase
  10.1103/physrevd.102.076014} {\bibfield  {journal} {\bibinfo  {journal}
  {Physical Review D}\ }\textbf {\bibinfo {volume} {102}},\ \bibinfo {pages}
  {076014} (\bibinfo {year} {2020})}\BibitemShut {NoStop}%
\end{thebibliography}

\providecommand{\noopsort}[1]{}\providecommand{\singleletter}[1]{#1}%

\end{document}